\documentclass{article}
\usepackage{latexsym}
\title{\vglue4.6truecm
Clifford spinors and the relativistic point particle}
\author{Kaare Borchsenius
\thanks{Bollerisvej 8, 3782 Klemensker, Denmark,
e-mail: bdge@post5.tele.dk}}
\date{}
\begin{document}
\maketitle
\begin{abstract}
We examine the structure of the Clifford algebra associated with a Hermitian
bilinear form and apply the result to a dynamical model of the relativistic
point particle. The dynamics of the particle is described by a Dirac spinor with
components in a Clifford algebra. This spinor determines, through the Clifford
algebra, both the space-time coordinates and their conjugate momenta and
satisfies a first order equation of motion which leads to the usual space-time
canonical equations of motion. The constraints appear as the equations of motion
for the einbein and spin connection which are needed to ensure the
reparametrization invariance and local Lorentz invariance of the action. 
\end{abstract}
\pagebreak
\section{Introduction}
Spinors are usually associated with field theories for Fermions. Since, however,
they are simply the `building blocks' of the Lorentz group, there is no reason
why other field theories could not, in principle, result from the dynamics of
underlying spinors. In~\cite{borch4} we demonstrated an example of this kind in
the simplest possible case, that of the relativistic point particle. We examined
a dynamical model in which the quantized canonical space-time coordinates $X$
of the particle were resolved into two-component spinors $C$ with components in
a Clifford algebra
\begin{equation} 
X^{A\dot{B}}_{ab} = \{ C^{A}_{a},C^{*\dot{B}}_{b} \}, \ \ \ \ \{
C^{A}_{a},C^{B}_{b} \}=0 ,\ \ \ \ X^{A\dot{B}}_{ab} \stackrel{\mathrm{def}}{=}
\sigma_{\mu}^{A\dot{B}}X^{\mu}_{ab}
\label{a1}
\end{equation}
where $X^{\mu}_{ab}$ transform under a Lorentz transformation in the index $\mu$
and under a unitary change of basis in Hilbert space in the indices $a$ and
$b$. The model also addressed the more general question as to whether space-time
has a substructure which relates the complex properties of the Lorentz group
and quantum mechanics. Spinor substructures of space-time have been discussed in
Schwartz and Van Nieuwenhuizen~\cite{schwartz} and in
Borchsenius~\cite{borch1,borch2,borch3} 
 
In this paper we examine the structure of the Clifford algebra which is
associated with a Hermitian bilinear form, and apply the result to a dynamical
model of the relativistic point particle. We find that the structure of the
generating space, taken together with the need to accommodate parity, points to
a Dirac spinor with Clifford components as the primary dynamical variable. This
Dirac spinor determines through the Clifford algebra both the space-time
coordinates and conjugate momenta of the particle and satisfies a first order
equation of motion which generates the usual space-time canonical equations of
motion. The fact that the momenta are no longer defined in terms of the
coordinates and their derivatives means that the problem in constrained dynamics
of eliminating the derivatives of the coordinates in favour of the momenta does
not arise. The constraints appear instead as the equations of motion for the
einbein and spin connection which are needed to ensure the reparametrization
invariance and local Lorentz invariance of the action.     
\section{Clifford algebras and Hermitian bilinear\\forms}
We recall that a real Clifford algebra arises naturally as the `square root' of
a real quadratic form $Q$ on a real linear space $V$ 
\begin{equation} 
v^{2}=Q(v) , \  v \in  V 
\label{b1} 
\end{equation} 
where $Q$ can have any signature $s$. In case $Q$ is degenerate, the algebra
contains Grassmann elements. On expanding $v$ on an orthogonal basis $e_{i}$ of
$V$, it follows that (\ref{b1}) is satisfied if $e_{i}$ satisfies the generating
algebra
\begin{equation} 
\frac{1}{2}\{e_{i},e_{j}\} = \delta _{ij}Q(e_{i})
\label{b2} 
\end{equation} 
To relate Hermitian bilinear forms on a complex linear space $V$ to real
Clifford algebras, we need to consider vectors in $V\oplus V$ which we shall
write as $(\begin{array}{c}u\\v \end{array}), u,v \in V$. Then the following
proposition holds:
\begin{quote}
\emph{Let $B$ be a Hermitian bilinear form of signature $s$ on an
$n$-di\-men\-sion\-al complex linear space $V$ with complex involution $\ast $.
Then there exists a $2n$-dimensional Clifford algebra of signature $2s$ for $V
\oplus V$ so that
\begin{equation} 
B(u,v) = \left \{ \left( \begin{array}{c}u\\0 \end{array} \right),\left(
\begin{array}{c}0\\v^{\ast } \end{array} \right) \right \}, \ \ \ \left \{
\left( \begin{array}{c}u\\0 \end{array} \right),\left(
\begin{array}{c}v\\0 \end{array} \right) \right \}=0
\label{b11}
\end{equation} }
\end{quote}
\emph{Proof} The involution $\ast $ of $V$ induces the complex involution
\begin{equation} 
\left( \begin{array}{c}u\\v \end{array} \right)^{\ast }
\stackrel{\mathrm{def}}{=} \left( \begin{array}{c}v^{\ast }\\u^{\ast }
\end{array} \right)
\label{b3} 
\end{equation}
of $V\oplus V$. Let $f_{i}$ be an orthogonal normalized basis for V
\begin{equation} 
B(f_{i},f_{j}) = \delta _{ij}g_{i}, \  g_{i} \in \{-1,0,1 \}
\label{b4} 
\end{equation}
Then $( \begin{array}{c}f_{i}\\0 \end{array} ) , (
\begin{array}{c}0\\f_{i}^{\ast }
\end{array} )$ is a basis for $V\oplus V$. It follows that the $2n$-dimensional
real subspace $W$ of  vectors $w = w^{\ast }, w \in  V\oplus V $ has the basis
\begin{equation} 
a_{j} = \frac{1}{2} \left( \begin{array}{c}f_{j}\\0 \end{array} \right)+
\frac{1}{2} \left( \begin{array}{c}0\\f_{j}^{\ast } \end{array} \right), \ \    
b_{j} = -\frac{i}{2} \left( \begin{array}{c}f_{j}\\0 \end{array} \right)+
\frac{i}{2} \left( \begin{array}{c}0\\f_{j}^{\ast } \end{array} \right) 
\label{b5} 
\end{equation}
For this space there exists a $2n$-dimensional real Clifford algebra of
signature $2s$ with the generating algebra
\begin{equation} 
\{a_{i},a_{j}\}=\{b_{i},b_{j}\}=\frac{1}{2}\delta _{ij}g_{j}, \ \ \ \ \
\{a_{i},b_{j}\}=0
\label{b6}
\end{equation}
Since $a_{i},b_{i}$ also form a basis for the whole of $V\oplus V$, the
Clifford product can be extended from the real subspace $W$ to the whole of
$V\oplus V$. This permits us to write the generating algebra (\ref{b6}) in the
equivalent form 
\begin{equation} 
\left \{ \left( \begin{array}{c}f_{i}\\0 \end{array} \right),\left(
\begin{array}{c}0\\f_{j}^{\ast } \end{array} \right) \right \}=\delta
_{ij}g_{j}, \ \ \ \ \  \left \{ \left( \begin{array}{c}f_{i}\\0 \end{array}
\right),\left(
\begin{array}{c}f_{j}\\0 \end{array} \right) \right \}=0
\label{b7}
\end{equation}
and (\ref{b11}) follows by expanding $u$ and $v$ on $f_{i}$.$\Box$

In matrix language, this proposition implies that any $n\times n$ Hermitian
matrix $H_{ij}$ of signature $s$ can be expressed in terms of the elements
$v_{i}$ of a $2n$-dimensional Clifford algebra
\begin{equation} 
H_{ij}=\{v_{i},v_{j}^{\ast }\}, \ \ \ \ \{v_{i},v_{j}\}=0
\label{b8}
\end{equation}
Applying this to the Hermitian spinor components of the space-time coordinates 
\begin{equation} 
x^{A\dot{B}} \stackrel{\mathrm{def}}{=} \sigma _{\mu}^{A\dot{B}}x^{\mu}
\label{b9}
\end{equation}
where $\sigma^{\mu} $ are the four Hermitian Pauli matrices, we obtain
\begin{equation} 
x^{A \dot{B}}= \left \{ \left( \begin{array}{c}c^{A}\\0 \end{array}
\right),\left( \begin{array}{c}0\\c^{\ast \dot{B} } \end{array} \right) \right
\}, \ \ \ \  \left \{ \left( \begin{array}{c}c^{A}\\0 \end{array} \right),\left(
\begin{array}{c}c^{B}\\o \end{array} \right) \right \}=0 
\label{b10}
\end{equation}
which was the basis for the discussion in ~\cite{borch4}.
\section{Dirac spinors with Clifford components}
In ~\cite{borch4} we started out with the single Weyl spinor $c^{A}$ in
(\ref{b10}) as the primary dynamical variable of the point particle. There are
two objections to this. First, the Clifford product is defined in $V\oplus V$;
so there is no formal reason to restrict the components of the primary variables
to $V$. Secondly, there is no simple way of representing space and time
inversion with only a single Weyl spinor. Both objections are met by taking
instead as basic dynamical variable the Dirac spinor
\begin{equation} 
\psi = \left( \begin{array}{c} \psi _{l}^{A}\\ \psi _{r \dot{B}}
\end{array} \right)
\label{c1}
\end{equation}
with left- and right-handed Weyl components $\psi _{l}$ and $\psi _{r}$. For
each value of $(A,B)$, $\psi $ is an element of $V\oplus V$. Whereas the Weyl
spinor in (\ref{b10}) determined only the space-time coordinates of the
particle, the Dirac spinor (\ref{c1}) will, as we shall see, determine both the
coordinates and their conjugate momenta.

To construct an action we must be able to form Lorentz invariant commutators
between Dirac spinors. From the Clifford algebra it follows that
\begin{equation}
\left \{ \left( \begin{array}{c}u\\0 \end{array}
\right),\left( \begin{array}{c}v\\0 \end{array} \right) \right
\}= \left \{ \left( \begin{array}{c}0\\u \end{array}
\right),\left( \begin{array}{c}0\\v \end{array} \right) \right
\}=0 
\label{c2}
\end{equation}
Hence, for the Dirac spinors $ \psi = ( \begin{array}{c} \psi _{l}\\ \psi _{r}
\end{array} )$ and $\chi  = ( \begin{array}{c} \chi  _{l}\\ \chi _{r}
\end{array})$
we obtain the commutator
\begin{equation} 
\{ \psi , \chi \}= \left \{ \left( \begin{array}{c} \psi_{l} \\ 0  \end{array}
\right),  \left( \begin{array}{c} 0\\ \chi_{r}  \end{array} \right)
\right \} + \left \{ \left( \begin{array}{c} 0\\\psi_{r}  \end{array} \right),
\left( \begin{array}{c} \chi_{l}\\0 \  \end{array} \right) \right \}
\label{c3}
\end{equation}
We observe that the Clifford product couples the Weyl spinors in the opposite
way of an ordinary contraction. It is therefore convenient to define the bracket
\begin{equation} 
\{ \psi, \chi \}_{_{\sim }}  \stackrel{ \mathrm{def}}{=}  \{ \psi ,
\stackrel{\sim}{ \chi} \} , \ \ \ \  \stackrel{\sim}{ \left( \begin{array}{c}
\chi_{l}\\
\chi_{r} \end{array} \right)}  \stackrel{ \mathrm{def}}{=} \left(
\begin{array}{c} \chi_{r}\\ \chi_{l}  \end{array} \right)   
\label{c4}
\end{equation}
The Dirac conjugate is defined in the usual way but with $\ast $ taking the
place of ordinary complex conjugation of the Weyl spinors 
\begin{equation} 
\bar{\psi } \stackrel{ \mathrm{def}}{=} (\psi _{l}^{\ast} ,\psi
_{r}^{\ast}) \gamma ^{0}
\label{c5}
\end{equation}
In the Weyl representation the Lorentz scalar corresponding to $\bar{\psi }\chi$
is therefore
\begin{equation} 
\{ \bar{\psi} ,\chi \}_{_{\sim }} = \left \{ \left( \begin{array}{c}
\psi_{rB}^{\ast } \\ 0 \end{array} \right),  \left( \begin{array}{c} 0\\
\chi_{l}^{B}  \end{array} \right) \right \} +  \left \{ \left( \begin{array}{c}
0\\\psi_{l}^{\ast \dot{A} }  \end{array} \right),
\left( \begin{array}{c} \chi_{r \dot{A}}\\0 \  \end{array} \right) \right \}
\label{c6}
\end{equation}

The basic Dirac spinor $\psi $ determines the complex four-vector 
\begin{equation} 
a^{\mu } = \{ \bar{\psi }_{c},\gamma ^{\mu }\psi \}_{_{ \sim }} , \ \ \ \  \psi
_{c}  \stackrel{ \mathrm{def}}{=} C\bar{\psi }^{T} 
\label{c7}
\end{equation}
which, since both $\psi $ and its charge conjugate $\psi _{c}$ transform like
Dirac spinors, will transform like a vector under space and time inversion. So
will its real and imaginary parts:
\begin{eqnarray} 
a^{\mu } = x^{\mu } + ip^{\mu }
\label{c8}
\\ x^{\mu } \stackrel{ \mathrm{def}}{=} \frac{1}{2}\{ \bar{\psi
}_{c},\gamma ^{\mu }\psi \}_{_{ \sim}
}+\frac{1}{2}\{ \bar{\psi },\gamma ^{\mu }\psi_{c} \}_{_{ \sim }}
\label{c9}
\\ p^{\mu } \stackrel{ \mathrm{def}}{=}  -\frac{i}{2} \{ \bar{\psi
}_{c},\gamma ^{\mu }\psi \}_{_{ \sim }}+\frac{i}{2} \{ \bar{\psi },\gamma ^{\mu
}\psi_{c} \}_{_{ \sim }}
\label{c10}
\end{eqnarray}
To show that $x^{\mu }$ and $p^{\mu }$ are independent vectors we express them
in terms of the Weyl components of $\psi $ using (\ref{c7})
\begin{equation} 
\sigma_{\mu}^{A\dot{B}}(x^{\mu}+ip^{\mu })= 4i\left \{ \left(
\begin{array}{c}\psi _{l}^{A}\\0 \end{array}
\right),\left( \begin{array}{c}0\\\psi _{r}^{\dot{B} } \end{array} \right)
\right\}    
\label{c11}
\end{equation}
In ~\cite{borch4} we showed that different eigenvalues of the quantized 
space-time coordinates $X$ corresponded to mutually anticommuting Weyl spinors.
Since $X$ has a continuous spectrum it follows that any two four-vectors can be
expressed in the form (\ref{b10}) with  mutually anticommuting Weyl spinors.
Using this property it follows immediately that (\ref{c11}) can be solved with
respect to $\psi _{l}$ and $\psi _{r}$ for arbitrary values of $x$ and $p$.
\section{Equations of motion}
The dynamics of the relativistic point particle will be described by the Dirac
spinor $\psi $ as a function of a parameter-time $\tau $. We consider the action
\begin{equation} 
\int_{}^{}{\left\{ \bar{\psi },i\left( \frac{d}{d\tau}+\Gamma(\tau )\right)\psi 
\right\}_{_{\sim}}+c.c.-e(\tau )H(x,p)\,d\tau }
\label{d1}
\end{equation}
which is invariant under a reparametrization $\tau \to \tau'$ of the world line
with the einbein $e(\tau)$ and spin connection $\Gamma(\tau )$ transforming as
\begin{equation}
e'(\tau' )= e(\tau )\frac{d\tau }{d\tau' }, \ \ \ \ \Gamma '(\tau ')=\Gamma
(\tau )\frac{d\tau }{d\tau' }
\label{d12}
\end{equation}
and under a local Lorentz transformation $\psi \to  S(\tau )\psi$ with the spin
connection $\Gamma (\tau )$ transforming as
\begin{equation}
\Gamma (\tau )\to S\Gamma (\tau )S^{-1}-\frac{dS}{d\tau }S^{-1}, \ \ \ \Gamma
(\tau )=i\Gamma _{\mu \nu }(\tau )\sigma ^{\mu \nu }, \ \ \ \sigma ^{\mu \nu
}\stackrel{ \mathrm{def}}{=} \frac{i}{2}[\gamma^{\mu }, \gamma^{\nu }] 
\label{d13}
\end{equation}
The equations of motion for $e(\tau )$ and $\Gamma (\tau )$ give the constraints
\begin{eqnarray}
H(x,p)=0 
\label{d14} \\
\{\bar{\psi },\sigma ^{\mu \nu }\psi\}_{_{\sim }} = 0 
\label{d15}
\end{eqnarray}
Because the parameter-space is one-dimensional we can choose the gauge
\begin{equation}
e(\tau )=1, \ \ \ \ \Gamma (\tau )=0
\label{d16}
\end{equation}
The equations of motion for $\psi $ are obtained by varying the action
(\ref{d13}) with respect to $\bar{\psi }$. The contributions from $x$ and $p$
come from $\bar{\psi }$ and $\psi _{c}$ in the second terms in (\ref{c9}) and
(\ref{c10}), and we obtain     
\begin{equation} 
\frac{d\psi}{d\tau} = \frac{1}{2}\left(\frac{\partial H}{\partial p^{\mu }} - i
\frac{\partial H}{\partial x^{\mu }}\right) \gamma ^{\mu }\psi _{c}
\label{d2}
\end{equation}
When applied to (\ref{c7}) this gives the equations of motion for $a^{\mu }$ 
\begin{equation} 
\frac{da^{\mu }}{d\tau} = \left(\frac{\partial H}{\partial p^{\nu  }} - i
\frac{\partial H}{\partial x^{\nu  }}\right) \{\bar{\psi },\gamma^{\nu }
\gamma^{\mu }\psi  \}_{_{\sim }} 
\label{d3}
\end{equation}
which reduce to 
\begin{equation} 
\frac{da^{\mu }}{d\tau} = \left(\frac{\partial H}{\partial p_{\mu }  } - i
\frac{\partial H}{\partial x_{\mu   }}\right) \{\bar{\psi },\psi  \}_{_{\sim }} 
\label{d9}
\end{equation}
by use of the constraint (\ref{d15}). For non-vanishing $\{\bar{\psi },\psi
\}_{_{\sim }}$ we can reparametrize these equations with the new parameter
$\bar{\tau }$
\begin{equation} 
\frac{d\bar{\tau }}{d\tau } = \{\bar{\psi },\psi  \}_{_{\sim }} 
\label{d10}
\end{equation}
and split them into real and imaginary parts to obtain the canonical equations
of motion
\begin{equation} 
\frac{dx^{\mu }}{d\bar{\tau}}=\frac{\partial H}{\partial p_{\mu   }} , \ \ \ \
\frac{dp^{\mu }}{d\bar{\tau}}= -\frac{\partial H}{\partial x_{\mu  }}
\label{d11}
\end{equation}
This leads to the interpretation of $x$, $p$ and $\bar{\tau }$ as the
coordinates, conjugate momenta and proper-time of the particle. 
\section{The free particle}
The free particle is described by the Hamiltonian
\begin{equation} 
H(x,p)=\frac{1}{2m}\left(p_{\mu }p^{\mu }- m^{2}\right)
\label{e1}
\end{equation}
with the constraint 
\begin{equation}
p_{\mu }p^{\mu }= m^{2}
\label{e2}
\end{equation}
The equations of motion (\ref{d2}) become
\begin{equation}
\frac{d\psi}{d\tau} = \frac{p_{\mu }}{2m}\gamma ^{\mu } \psi _{c} 
\label{e3}
\end{equation}
From (\ref{d9}) it follows that $p$ is constant. Using the constraint
(\ref{e2}) we obtain the complete solution to (\ref{e3}) 
\begin{equation}
\psi = \alpha e^{\frac{1}{2}\tau } + \beta e^{-\frac{1}{2}\tau }
\label{e4}
\end{equation}
where the Dirac spinors $\alpha $ and $\beta $ are constants of integration
which satisfy
\begin{equation}
\alpha = \frac{p_{\mu }}{m}\gamma ^{\mu }\alpha  _{c}, \ \ \ \ \beta  = -
\frac{p_{\mu }}{m}\gamma ^{\mu }\beta  _{c}
\label{e5}
\end{equation}
From (\ref{e5}) it follows that
\begin{equation}
\{\bar{\alpha },\beta \}_{_{\sim }} = -\{\bar{\beta  },\alpha  \}_{_{\sim }}
\label{e6}
\end{equation}
so that $\{\bar{\alpha },\beta \}_{_{\sim }}$ is imaginary. The constraint
(\ref{d15}) gives 
\begin{eqnarray}
\{\bar{\alpha },\sigma ^{\mu \nu }\alpha \}_{_{\sim }} = \{\bar{\beta  },\sigma
^{\mu \nu}\beta  \}_{_{\sim }} = 0
\label{e7}
\\
\{\bar{\alpha },\sigma ^{\mu \nu }\beta  \}_{_{\sim }} + \{\bar{\beta  },\sigma
^{\mu \nu}\alpha \}_{_{\sim }} = 0
\label{e8}
\end{eqnarray}
$a^{\mu}$ is obtained by inserting the expression (\ref{e4}) for $\psi $ into
(\ref{c7}) 
\begin{eqnarray}
a^{\mu}=
\frac{p_{\nu}}{m}\left(\{\bar{\alpha},\gamma^{\nu}\gamma^{\mu}\alpha\}_{_{\sim
}}e^{\tau } - \{\bar{\beta},\gamma^{\nu}\gamma^{\mu}\beta\}_{_{\sim}}e^{-
\tau}\right) \nonumber \\
 \mbox{}+ \frac{p_{\nu}}{m}\left(\{\bar{\alpha},\gamma^{\nu }\gamma^{\mu
}\beta\}_{_{\sim }}- 
 \{\bar{\beta},\gamma^{\nu }\gamma^{\mu }\alpha \}_{_{\sim }}\right) 
\label{e9}
\end{eqnarray}
To obtain the coordinates and momenta we split $a^{\mu }$ into its real and
imaginary parts by applying conditions (\ref{e6}), (\ref{e7}) and (\ref{e8}) 
\begin{eqnarray}
p^{\mu } = -\frac{2i}{m}\{\bar{\alpha },\beta \}_{_{\sim }}p^{\mu }
\label{e10}
\\
x^{\mu }= \frac{p^{\mu }}{m}\left(\{\bar{\alpha },\alpha \}_{_{\sim }}e^{\tau }-
\{\bar{\beta},\beta \}_{_{\sim }}e^{-\tau }\right) - 2i\{\bar{\alpha },\sigma
^{\nu \mu}\beta \}_{_{\sim }}p_{\nu }
\label{e11}
\end{eqnarray}
(\ref{e10}) simply adjusts the value of $\{\bar{\alpha },\beta \}_{_{\sim }}$ to
be consistent with the constant value of $p^{\mu }$. The proper-time is found
by integrating (\ref{d10})
\begin{equation}
\frac{d\bar{\tau }}{d\tau } = \{\bar{\psi },\psi  \}_{_{\sim }}= \{\bar{\alpha
},\alpha \}_{_{\sim }}e^{\tau }+\{\bar{\beta  },\beta  \}_{_{\sim }}e^{-\tau }
\label{e13}
\end{equation}
to give
\begin{equation}
\bar{\tau }=\{\bar{\alpha },\alpha \}_{_{\sim }}e^{\tau }-\{\bar{\beta  },\beta 
\}_{_{\sim }}e^{-\tau} + \tau _{0}
\label{e14}
\end{equation}
The expression for $x$ in terms of proper-time is obtained from (\ref{e11}) and
(\ref{e14}):
\begin{equation}
x^{\mu }(\bar{\tau })= \frac{p^{\mu }}{m}\bar{\tau }+x^{\mu }(0),\ \ \ \ x^{\mu
}(0)\stackrel{ \mathrm{def}}{=} -\frac{p^{\mu }}{m}\tau _{0}-2i\{\bar{\alpha
},\sigma ^{\nu \mu }\beta\}_{_{\sim }}p_{\nu }
\label{e15} 
\end{equation}
The solutions fall into two classes depending on the relative signs of
$\{\bar{\alpha },\alpha \}_{_{\sim }}$ and $\{\bar{\beta  },\beta  \}_{_{\sim
}}$. When they have the same signs a complete solution to the equations of
motion for $\psi$ generates a complete solution to the space-time equations of
motion. When they have opposite signs the space-time solution has an endpoint in
proper-time and the complete solutions to the equations of motion for $\psi$
are
double coverings of incomplete solutions to the space-time equations. Before
rejecting this last type of solution as unphysical the possibility should be
kept in mind that in a realistic theory a common starting point in proper-time
could have a cosmological interpretation. The double covering property also
provides a simple local interpretation of the quantum interference between
alternative space-time paths ~\cite{borch4}. 
\section{Conclusion}
The model of the relativistic point particle discussed in the foregoing
suggests that the concept of canonically conjugate variables has an algebraic
origin in the relationship between Hermitian bilinear forms and Clifford
algebras. The $V\oplus V$ structure of the generating space of the Clifford
algebra lends itself naturally to a representation of parity in terms of a Dirac
spinor with Clifford components. From this spinor the space-time coordinates
and momenta emerge in a unified manner, and the problem in constrained dynamics
of eliminating the derivatives of the coordinates in favour of the momenta does
not arise. 

It is customary to obtain a many-particle theory by applying the coordinate
representation to the constraints. However, since the space-time coordinates are
no longer primary variables, such a procedure would only be an approximation.
Another option which naturally suggests itself in the present context is to
increase the dimension of the parameter-space by adding a world-sheet spinor
index to the world spinor index of $\psi $. It remains to be seen which of the
features of the point particle model discussed in the foregoing can be
carried over into such a higher dimensional theory.        

\end{document}